# Pick-up and drop transfer of diamond nanosheets

**V Seshan[1,2,*], J O Island[1,*], R van Leeuwen[1], W J Venstra[1], B H Schneider[1], S D Janssens[3,4], K Haenen[3,4], E J R Sudhölter[2], L C P M de Smet[2], H S J van der Zant[1], G A Steele[1] and A Castellanos-Gomez[1, +]**

[1] Kavli Institute of Nanoscience, Delft University of Technology, Lorentzweg 1, 2628 CJ Delft, The Netherlands
[2] Department of Chemical Engineering, Delft University of Technology, Julianalaan 136, 2628 BL Delft, The Netherlands
[3] Institute for Materials Research (IMO), Hasselt University, Wetenschapspark 1, BE-3590 Diepenbeek, Belgium
[4] IMOMEC, IMEC vzw, Wetenschapspark 1, BE-3590, Diepenbeek, Belgium
[+] Present Address: Instituto Madrileno de Estudios Avanzados en Nanociencia (IMDEA - Nanociencia), 28049 Madrid (Spain)

[+] E-mail: andres.castellanos@imdea.org

[*]These authors contributed equally to this work.

**Abstract**
Nanocrystalline diamond (NCD) is a promising material for electronic and mechanical micro- and nanodevices. Here we introduce a versatile pick-up and drop technique that makes it possible to investigate the electrical, optical and mechanical properties of as-grown NCD films. Using this technique, NCD nanosheets, as thin as 55 nm, can be picked-up from a growth substrate and positioned on another substrate. As a proof of concept, electronic devices and mechanical resonators are fabricated and their properties are characterized. In addition, the versatility of the method is further explored by transferring NCD nanosheets onto an optical fibre, which allows measuring its optical absorption. Finally, we show that NCD nanosheets can also be transferred onto 2D crystals, such as $MoS_2$, to fabricate heterostructures. Pick-up and drop transfer enables the fabrication of a variety of NCD-based devices without requiring lithography or wet processing.

## 1. Introduction

Diamond is a promising material that offers excellent mechanical, thermal and optical properties, such as a high Young's modulus, a high thermal conductivity and broadband transparency [1]. Pristine or undoped diamond is electrically insulating, however it can be turned into a semiconductor or superconductor by adding dopants such as boron, allowing diamond-based structures to be incorporated as functional components in electronic circuits [2, 3]. In addition to being biologically compatible and inert to harsh environments, diamond can be functionalized chemically in various ways, as to enable bio-based electronic sensing [4]. Moreover, diamond films, such as the nanocrystalline types, can be grown over large areas on non-diamond substrates, which significantly reduces the production costs [5]. The attractive properties and robustness of diamond opens new avenues for interesting applications, which include high-quality nanophotonic circuits, ultra-sensitive force and mass transducers operating under extreme conditions, nanoelectromechanical contact switches and bio-nanomechanical devices [6-10]. For many of these applications, diamond structures are typically fabricated using a top-down approach involving lithography, high-temperature annealing, etching and wet-chemical processing [11]. Creating diamond-based devices with more sensitive materials such as two-dimensional (2D) atomic crystals is challenging because diamond growth requires high temperatures in a hydrogen rich atmosphere [12, 13]. Therefore, a technique that allows the integration of a diamond film into a pre-fabricated device is desirable and opens doors to more complex device architectures exploiting the properties of diamond films.

Here, we propose a versatile pick-up and drop method using visco-elastic stamps to transfer nanocrystalline diamond (NCD) nanosheets. The process is based on an all-dry technique that allows transfer of the NCD nanosheets, doped or undoped, from one substrate to another. The dry transfer technique has been used extensively to deposit 2D materials with thicknesses of a few nanometers onto silicon-based substrates; here we use it to transfer NCD films onto a variety of substrates including metal electrodes, optical fibers and $MoS_2$ crystals [14, 15]. Using this technique, we demonstrate the fabrication of NCD-based electronic devices and mechanical resonators and characterize their properties. Additionally, an NCD nanosheet is transferred onto the center of an optical fiber and on top of $MoS_2$ flakes to demonstrate that transfer is feasible on a variety of substrates. Since there is no wet-chemical processing step involved in the pick-up and drop technique, it can be efficiently used as a final fabrication step to integrate NCD nanosheets in more complex devices or in pre-fabricated circuits.

## 2. Experimental section

*Growth of NCD nanosheets*: To grow the NCD nanosheets, we start with quartz substrates that are seeded with diamond nanoparticles [16]. The NCD films are deposited using a microwave plasma-enhanced chemical vapour deposition (CVD) process, using a conventional $H_2/CH_4$ plasma with methane concentration of ~4 % (v/v). The microwave power is maintained at 3500 W, the substrate temperature at 510-560 °C and the process pressure at 33-40 mbar. For obtaining the boron-doped NCD films used in electrical measurements, trimethylboron gas (boron to carbon concentration ratio of ~3000 ppm) is introduced in addition to $H_2/CH_4$ gases during the CVD growth process. The thickness of the NCD film is monitored in-situ using a laser interferometer. It is important to note that the CVD process is stopped when the thickness of the NCD film on quartz substrate reaches ~180 nm. During the growth, a mismatch in the thermal expansion coefficient of the quartz substrate and the NCD film results in the accumulation of stress between the two materials.

The conditions were purposefully chosen so that at a thickness of ~180 nm, this stress is sufficient to crack the film and to delaminate it from the quartz surface, forming numerous nanosheets. These nanosheets, typically 50 μm × 50 μm in size, remain weakly adhered to the quartz surface.

*Thinning of NCD nanosheets*: The NCD nanosheets are thinned down using an oxygen ($O_2$) reactive ion etching (RIE). A Leybold Heraeus system is used, with a DC bias voltage of approx. –413 V, $O_2$ gas flow of 30 ml/min and pressure of ~20.7 μbar for ~10 min. This results in an etching rate of ~15 nm/min. The NCD nanosheets that are ~185 nm thick initially are reduced to ~55 nm. The whole transfer method including RIE is all-dry and no solvent is used throughout the manufacturing process in contrast with typical wet-etching techniques [11].

*Thickness determination of NCD nanosheets*: To determine the thickness of the NCD nanosheets before and after RIE processing, tapping-mode atomic force microscopy (AFM) is performed using a Digital Instruments D3100 AFM with a standard silicon cantilevers (spring constant of 40 N/m and tip curvature of <10 nm).

*Electrical transport measurements*: The devices are electrically characterized using a two-terminal voltage bias in a Lakeshore Cryogenics probe station at room temperature and in ambient conditions.

*Mechanical motion detection*: The mechanical motion of the NCD nanosheet is measured using an optical interferometer, as described in more detail previously [15, 17]. Briefly, the setup consists of a Helium-Neon probing laser ($\lambda$ = 632.8 nm) focused on the suspended part of the nanosheet and a blue diode laser ($\lambda$ = 405 nm) with an optical output below 1 mW for photothermal excitation of the resonators. While the pre-patterned $SiO_2$ substrate with a 100 nm thick oxide square with a hole acts as the fixed mirror, the nanosheet acts as the semi-transparent moving mirror, thus forming an interferometer. On photothermal excitation, the motion of the NCD nanosheet changes the distance between the mirrors and via constructive or destructive interference modulates the reflected optical power. The intensity of the reflected optical signal is detected using a photodiode. All the measurements are carried out in vacuum (~$10^{-5}$ mbar) to reduce viscous air damping.

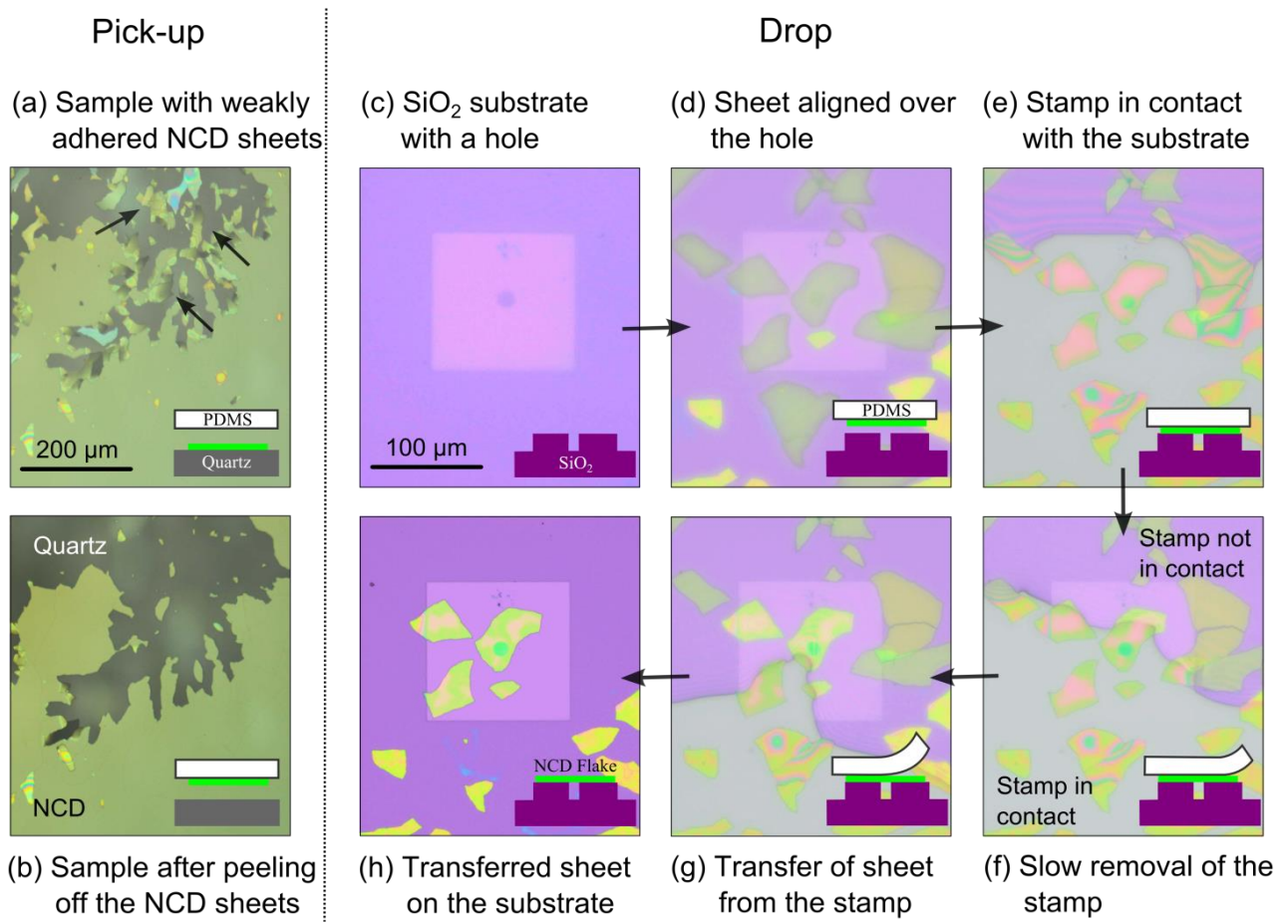

**Figure 1. Pick-up and drop transfer technique:** Optical microscopy images of the NCD nanosheets on a quartz substrate (a) before and (b) after stamping with a visco-elastic material. The arrows in panel (a) indicate the NCD nanosheets weakly adhered to the quartz substrate. (c-h) Series of optical microscopy images: (c) a pre-patterned SiO2 substrate with a 100 nm thick oxide square with a hole (diameter: 15 μm) in the middle; (d) the selected NCD nanosheet aligned over the hole on a SiO2 substrate using the micromanipulator; (e)-(g) the stamp brought in contact with the substrate and peeled-off using the micromanipulator; (h) the NCD nanosheet transferred over the hole on a SiO2 substrate. The insets show a schematic of the stamping process.

## 3. Results and discussion

Figure 1a shows an optical image of a quartz surface with a collection of delaminated NCD nanosheets. A transparent visco-elastic stamp (GelFilm® by GelPak), similar to those used in nano-imprinting, is brought into contact with the sample supporting the NCD nanosheets using a micromanipulator [14, 18]. Once in contact, due to the visco-elasticity, the stamp material can follow the topography of the NCD nanosheets that are weakly adhered to the substrate. When the stamp is rapidly (in one quick motion) peeled-off, NCD nanosheets are transferred from the (substrate) surface onto the stamp. Figure 1b shows an optical image of the same sample after peeling-off the stamp displaying release of nanosheets from the substrate surface. The NCD nanosheets can be picked-up from the substrate by the visco-elastic stamp and transferred to another surface. Figures 1c-f show an example of a transfer process where a 185 nm thick NCD nanosheet has been deposited on the center of a lithographically defined structure (a $100 \times 100$ μm$^2$ SiO$_2$ square with a 15 μm diameter hole in the center). The transfer process is carried out as follows using a method that was previously developed to transfer of 2D materials [14]. The stamp is mounted in a 3-axis micromanipulator with the surface containing the NCD nanosheets facing the substrate and aligned over the desired structure under a zoom lens (Figure 1d). By lowering the manipulator, the stamp is first brought into contact with the substrate and then pressed slightly against the substrate. This step is followed by a slow peel-off process (approx. 5-10 min) using the micromanipulator, which transfers the NCD nanosheet over the user-defined position. Figures 1e-h show the transfer process of the nanosheet from the visco-elastic stamp to the pre-patterned SiO$_2$ substrate. Figure 1h shows the resulting NCD nanosheet transferred on the hole at the center of the SiO$_2$ square. The whole manufacturing process of peeling off and transfer of nanosheets is accomplished in 10-15 min. The transparency of the visco-elastic stamp along with the sub-micron control of the placement by using an optical microscope allows precise positioning of the NCD nanosheet over the holes.

The pick-up and drop technique described above can be conveniently utilized to transfer/position NCD nanosheets with a sub-micron precision onto pre-patterned electrodes. We now turn our attention to the measurement of the electrical properties of transferred boron doped (B-NCD) nanosheets. Although NCD is intrinsically insulating, its electronic properties can be tailored to a great extent by doping it with boron atoms [1]. For example, B-NCD becomes conducting and with high enough doping it displays superconductivity at $T_c$ = 2.1 K [19]. This wide tunability of NCD (electronic) properties makes it very attractive for electronic applications and the possibility to transfer very thin doped-NCD films onto specific locations in a nanocircuit opens the door to the fabrication of complex device architectures. Furthermore, this technique could also offer a unique solution to repair micro-circuits by bridging the trenches or electrodes with B-NCD nanosheets. Figure 2a shows a scanning electron microscopy (SEM) image of a B-NCD nanosheet transferred between onto pre-patterned metallic electrodes (5 nm Ti/50 nm Au) on a Si/SiO$_2$ substrate. Two such devices with B-NCD nanosheets were fabricated and their electrical properties were characterized. Figure 2c shows the current *vs.* voltage characteristic of the devices measured at room temperature displaying two terminal resistances of 3.9 kΩ and 6.6 kΩ. The difference in the terminal resistances are attributed to the difference in the length and width of the electrodes. For the two fabricated devices, a resistivity of 6.0 Ω-cm is obtained, which agrees reasonably well with the reported value of 3.4 Ω-cm for a similar type of B-NCD films [20]. A zoom-in high angle SEM image of the nanosheet between the two electrodes is shown in Figure 2b. It is interesting to note that the nanosheet between the two electrodes is freely suspended. As many of the micro/nanoelectromechanical systems applications require structures to be free-standing and suspended from a substrate, our technique provides an all-dry (no solvent) and simple solution to fabricate suspended structures.

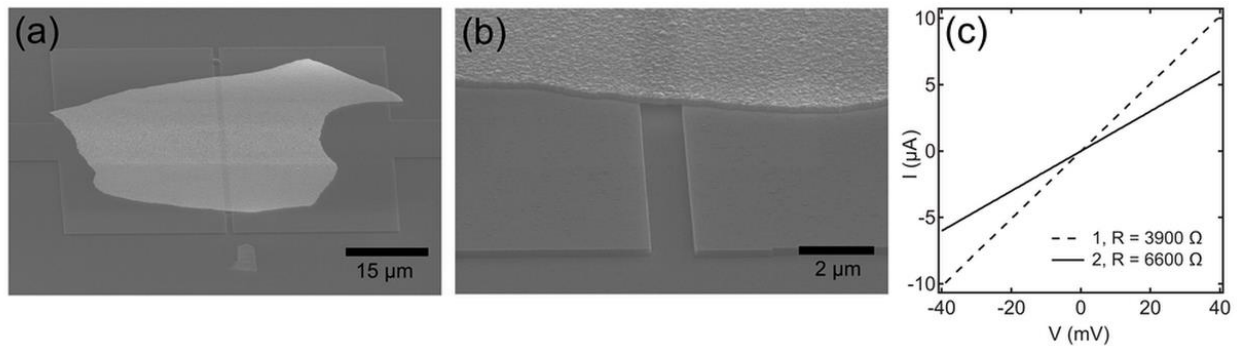

**Figure 2.** (a) Scanning electron microscopy image of the B-NCD nanosheet transferred onto Au electrodes and (b) corresponding zoom-in high angle image displaying the freestanding nature of the nanosheet. (c) Current *vs.* voltage characteristics of two different B-NCD devices.

As a proof of concept to fabricate freely suspended NCD nanosheets, mechanical drumhead resonators are fabricated (Figure 3). Five resonator devices with thicknesses of ~55 nm and ~185 nm and diameters from 6 μm to 15 μm were measured. The resonant motion of the NCD resonators are detected using an optical interferometer [15, 17]. Figure 3a shows the measured mechanical magnitude (circles) and phase (squares) spectra, with the corresponding optical microscopy image for the devices with the thickness of ~55 nm and (hole) diameters of 15 μm. The resonance frequencies and Q-factors are extracted by fitting the measured data to a damped-harmonic oscillator model, indicated by solid lines in Figure 3a. The resonance frequency of the devices is found to be in the range of 5 to 20 MHz (depending on thickness and diameter) with Q-factors between 40-155 (Figures 3a-d). For devices with identical thickness measured under the same experimental conditions, we observe a significant increase of the Q-factor with the resonator diameter. This suggests that the Q-factor is limited by clamping losses.

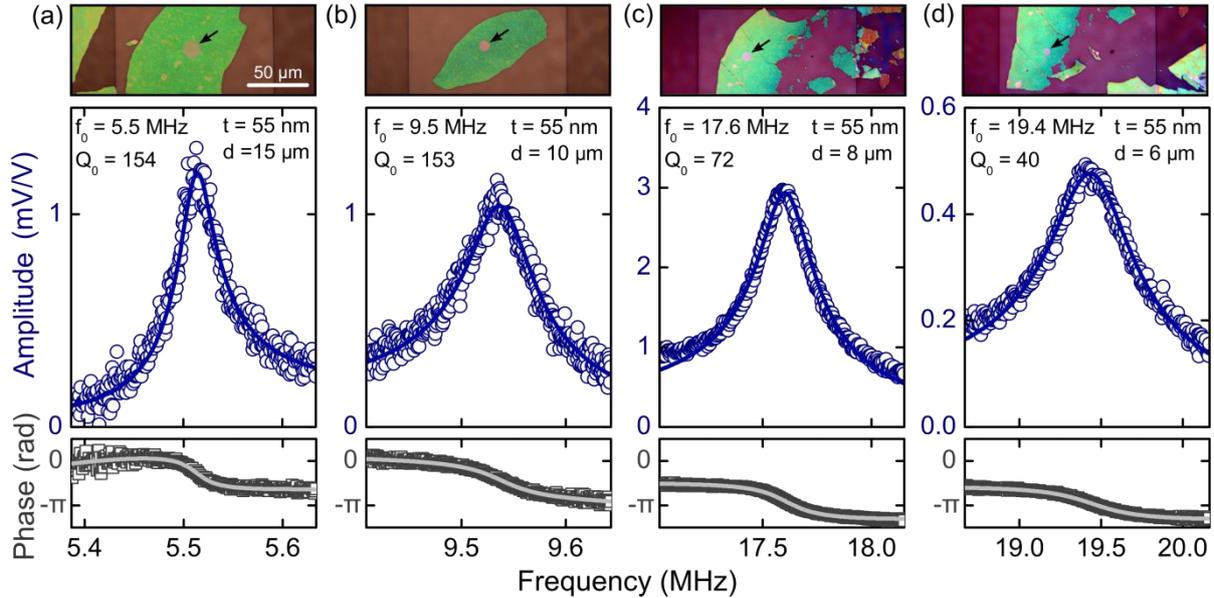

**Figure 3.** Measured normalized amplitude and phase response with corresponding optical image for the NCD mechanical resonator on a pre-patterned SiO2 substrate with a (hole) diameter of (a) 15 μm, (b) 10 μm, (c) 8 μm and (d) 6 μm. The arrow in the optical image indicates the location of the NCD mechanical resonators. The measured data (magnitude and phase) are fitted to a damped-driven harmonic oscillator model (solid lines) to obtain their resonance frequency and quality factor.

To characterize the obtained NCD resonators further, the resonance frequency ($f_0$) is measured as a function of the resonator thickness ($t$) and diameter ($d$). Figure 4a shows the measured resonance frequency versus the (resonator) thickness over the square of hole diameter for ~55 nm and ~185 nm thick NCD mechanical resonators. The uncertainty shown in the error bar in Figure 4a is due to the roughness of the NCD flake, which is approx. ± 15 nm (Figure 4b).

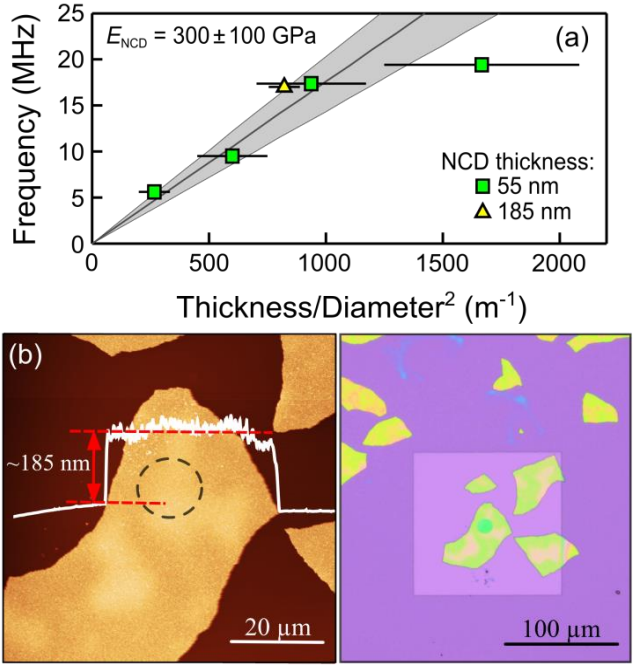

**Figure 4.** (a) Measured resonance frequency as a function of the (resonator) thickness and hole diameter. The solid area indicates the calculated resonance frequencies with $E = 300 \pm 100$ GPa, $\rho = 3500$ kg m$^{-3}$ and $v = 0.12$ (Equation 1). (b) Left panel shows the AFM image and the thickness profile of an NCD resonator/nanosheet. The thickness was found to be ~185 nm. The dotted circle in the AFM image represents the location of a pre-patterned hole (diameter: 15 μm) on the SiO2 substrate. Right panel shows the corresponding optical image of the NCD resonator suspended over the hole (diameter: 15 μm) on SiO2 substrate.

For a plate-like circular resonator clamped around its perimeter, the frequency is given by [21]:

$$f_0 = \frac{10.21}{\pi}\sqrt{\frac{E}{3\rho(1-v^2)}}\frac{t}{d^2} \quad [1]$$

where, $E$ is the Young's modulus, $\rho$ is the mass density and $v$ is the Poisson's ratio. For typical values for polycrystalline diamond of $E$ (≈304 GPa) [22], $\rho$ (3500 kgm$^{-3}$) [22] and $v$ (0.12) [23], the resonance frequencies calculated from Equation (1) are shown by the solid line in Figure 4a. The good correspondence between the measured data and this solid line indicates that the resonators indeed behave as circular plates whose dynamics is thus dominated by bending rigidity and not by their initial pre-tension. This is a desirable feature for nanomechanical resonators as the bending rigidity can be easily controlled by geometrical factors while the initial pre-tension typically varies from device to device. For a pre-tension dominated device, the second-order eigenmode is expected to be at 1.59 $f_0$ whereas for a bending rigidity dominated device, this value is expected to be around 2.08 $f_0$ [15, 21]. Therefore, to further confirm the plate-like behaviour of the NCD resonators, we measured the resonance frequencies of higher-order eigenmodes. Here we observe, for the second mode, resonance frequencies at 1.91 $f_0$ to 2.06 $f_0$, which is in good agreement with the expected value for a circular plate (Figure 5).

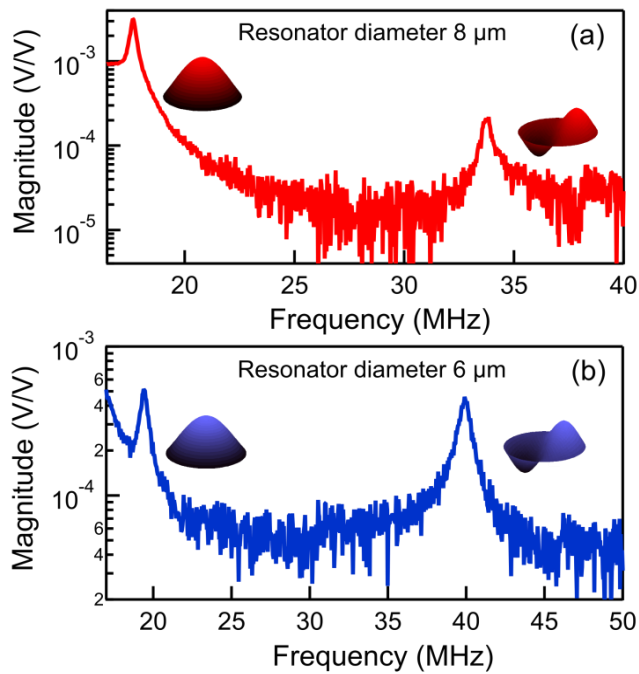

**Figure 5.** Resonance spectra displaying the fundamental and first higher-order modes of a 55 nm thick NCD resonator with a diameter of 8 µm (a) and 6 µm (b).

To further illustrate the versatility of our pick-up and drop transfer technique, a single NCD nanosheet (≈185 nm) is transferred onto the core of a multimode optical fiber. Figure 6 shows an optical image of the fiber (a) prior to and (b) after stamping the NCD nanosheet and (c) their corresponding optical transmission spectra. The transmittance ($T = I/I_0$), calculated from the transmission spectra, at 750 nm is ≈60 % which is comparable to reported values in literature for NCD films of similar thickness (≈62 %) [24]. This example demonstrates that this rather simple technique can be used to study the optical properties of NCD films and even to integrate these films as optically active components of an optical setup. The stamping technique can also be used to manufacture heterostructures through artificially stacking NCD nanosheets onto 2D materials such as $MoS_2$. Figures 6d-e display optical images of the $MoS_2$ flake on a $Si/SiO_2$ substrate followed by stacking of the NCD nanosheet on top of $MoS_2$ flake. The stamping method promises new applications and device components, such as electrodes, active materials, barrier materials and dielectric materials derived from stacking a wide variety of 2D materials in combination with NCD [25, 26].

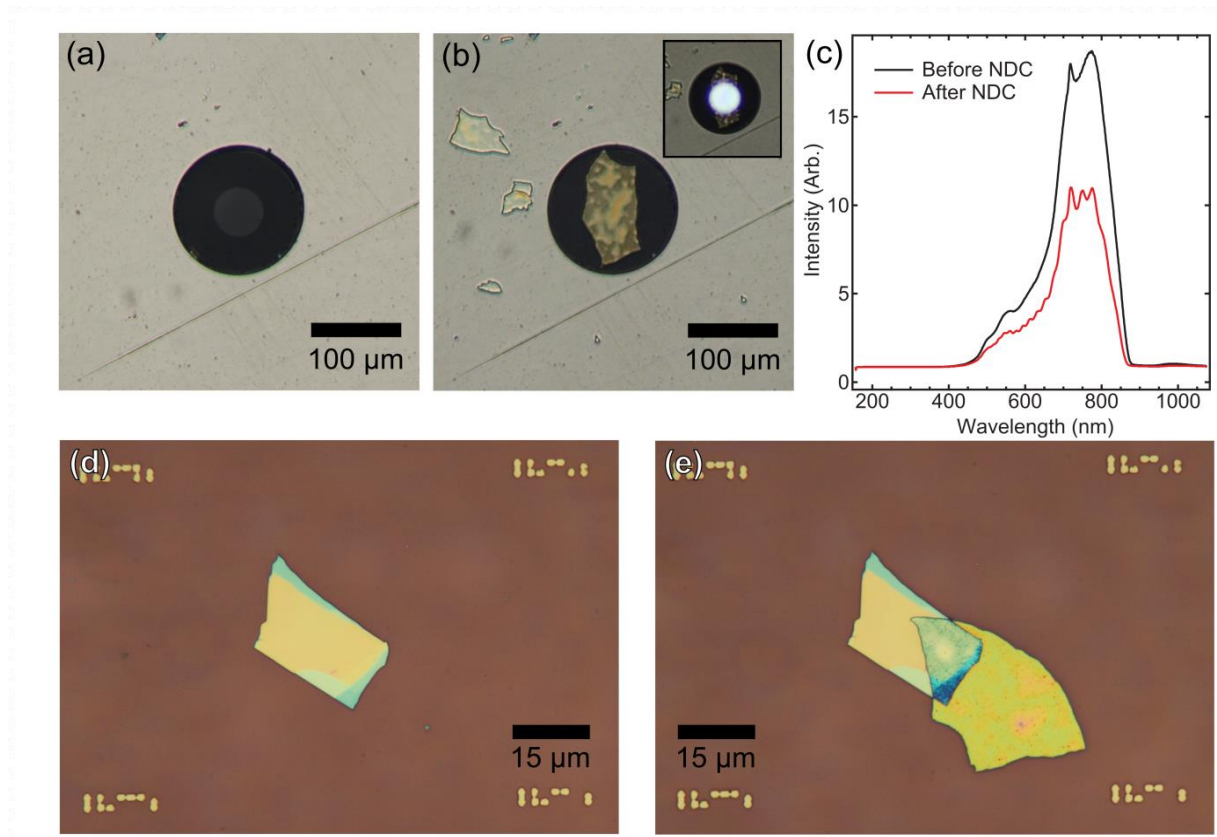

**Figure 6.** Optical microscopy images of an optical fiber (a) before and (b) after transferring a NCD nanosheet using a visco-elastic material. Inset shows the same transferred flake with the optical line illuminated. The corresponding optical transmission spectra measured using spectrometer is shown in panel (c). Fabrication of heterostructures based on artificial stacking: (d) Optical microscopy image of the $MoS_2$ flake on a silicon oxide substrate before and (e) after stacking the NCD nanosheet using our stamp transfer technique.

## 4. Conclusion

In summary, we demonstrate a versatile transfer technique based on an all-dry visco-elastic stamping to pick-up NCD films (as thin as 55 nm) from one substrate and to deposit them onto a specific location on another substrate. Additionally, we show that this relatively simple transfer technique can be employed to place NCD nanosheets directly onto pre-fabricated electrodes. As the NCD nanosheets can be transferred in the final step, it allows one to integrate NCD nanosheets in complex device architectures or in devices requiring very harsh fabrication processes or chemical treatments. Interestingly, the transferred NCD nanosheets can be freely-suspended without collapsing, making it possible to use this technique to fabricate NCD resonators. We characterize these resonators and measure frequencies of 5 to 20 MHz (depending on thickness and diameter) and Q-factors between 40-155. Finally, we demonstrate the flexibility of the pick-up and drop technique by transferring a NCD nanosheet onto the core of an optical fiber and on top of a few-layer $MoS_2$ flake, illustrating how this technique can be employed to transfer NCD nanosheets onto non-conventional substrates widening the range of applications of thin NCD films.

**Acknowledgements**

V Seshan and J O Island contributed equally to this work. The authors thank Shun Yanai (Delft University of Technology) for his help with the SEM and Milos Nesládek (Hasselt University) for insightful discussions. The authors would like to acknowledge the funding agencies: the Delft University of Technology, the Dutch organization for Fundamental Research on Matter (FOM), the Research Program G.0456.12N of the Research Foundation-Flanders (FWO), the FP7-Marie Curie Project PIEF-GA-2011-300802 ('STRENGTHNANO'), NanoNextNL (a micro and nanotechnology consortium of the Government of the Netherlands and 130 partners) and the European Union's Seventh Framework Programme (FP7/2007-2013) under Grant Agreement n° 318287, project LANDAUER.